\pdfoutput=1
\documentclass[journal]{IEEEtran}
\usepackage{booktabs} 
\usepackage{amsmath,amssymb,amsfonts}
\usepackage{algorithmic}
\usepackage{graphicx}
\usepackage{caption}
\usepackage[linesnumbered,ruled]{algorithm2e}
\usepackage{multirow}
\usepackage{subcaption} 
\ifCLASSINFOpdf
\else
\fi
\begin{document}

\title{Synchronous Multi-modal Semantic Communication System with Packet-level Coding}

\author{Yun Tian, Jingkai Ying, Zhijin Qin, ~\IEEEmembership{Senior Member,~IEEE}, Ye Jin, and Xiaoming Tao, ~\IEEEmembership{Senior Member,~IEEE}

\thanks{This work was supported in part by the National Natural Science Foundation of China (NSFC) under Grant 62293484 and 61925105. \textit{(Corresponding author: Zhijin Qin)}}
\thanks{
Yun Tian and Ye Jin are with the School of Electronics, Peking University, Beijing 100091, China. (email: tianyun@stu.pku.edu.cn; jinye@pku.edu.cn).}
\thanks{
Jingkai Ying, Zhijin Qin, and Xiaoming Tao are with the Department of Electronic Engineering, Tsinghua University, Beijing 100084, China, and with the State Key Laboratory of Space Network and Communications, Beijing, China, and also with Beijing National Research Center for Information Science and Technology, Beijing, China. (email: yjk23@mails.tsinghua.edu.cn; qinzhijin@tsinghua.edu.cn; taoxm@tsinghua.edu.cn).
}}

\maketitle

\begin{abstract}
Although the semantic communication with joint semantic-channel coding design has shown promising performance in transmitting data of different modalities over physical layer channels, the synchronization and packet-level forward error correction of multimodal semantics have not been well studied. Due to the independent design of semantic encoders, synchronizing multimodal features in both the semantic and time domains is a challenging problem. 
In this paper, we take the facial video and speech transmission as an example and propose a Synchronous Multimodal Semantic Communication System (SyncSC) with Packet-Level Coding. To achieve semantic and time synchronization, 3D Morphable Mode (3DMM) coefficients and text are transmitted as semantics, and we propose a semantic codec that achieves similar quality of reconstruction and synchronization with lower bandwidth, compared to traditional methods. To protect semantic packets under the erasure channel, we propose a packet-Level Forward Error Correction (FEC) method, called PacSC, that maintains a certain visual quality performance even at high packet loss rates. Particularly, for text packets, a text packet loss concealment module, called TextPC, based on Bidirectional Encoder Representations from Transformers (BERT) is proposed, which significantly improves the performance of traditional FEC methods. The simulation results show that our proposed SyncSC reduce transmission overhead and achieve high-quality synchronous transmission of video and speech over the packet loss network.
\end{abstract}


%
\IEEEpeerreviewmaketitle

\section{Introduction}

\IEEEPARstart{W}{ith} the explosion of intelligent applications and network traffic, conventional communication systems are facing the challenge of big data transmission. In recent years, deep learning enabled semantic communication systems have gathered attention \cite{aiempowered}. Compared to symbol transmission in traditional communication systems, 
semantic communication systems with focusing on transmitting the semantics of the data,  reduce efficiently bandwidth and improving communication efficiency \cite{qin2021semantic}. 
Despite these advantages from semantic coding, the synchronous transmission of multimodal data presents significant challenges.
During human-computer interaction, a user typically generates multimodal data simultaneously, such as video, audio, text, etc. Multimodal data from the same source show strong correlations and ambiguous segmentation \cite{liang2022foundations}. These multimodal data are highly aligned in both the semantic and time domains. 
Because highly compressed semantic features from semantic encoders designed independently lack alignment information, it is challenging to synchronize the transmitted data. 
Besides, when faced with packet loss network, semantic features are only received perfectly or lost. However, physical layer semantic communication systems have no higher-level joint designs. The semantics transmitted through existing network protocols are difficult to cope with changing network environments. Therefore, the semantic packet-level forward error correction is needed to be addressed.

\subsection{Piror Work}
Compared to conventional communication systems, semantic communication systems designed with Deep Neural Network (DNN) perform well for various type of information sources and tasks. According to the modality of the transmission source, current semantic communication systems can be divided into two categories, namely single-modal and multimodal systems. For single-modal sources or tasks such as text \cite{farsad2018deep,xie2021deep,jiang2022deep}, speech \cite{weng2021semantic}, 
image \cite{huang2022toward,kurka2020deepjscc,jankowski2020wireless}, and video \cite{wang2022wireless}, semantic communication systems extract and transmit tailored semantic features related to tasks at the receiver side. These works have reduced the transmission overhead for single-modal sources and shown notable performance in low signal-to-noise (SNR) regime.

For multimodal tasks or transmission, existing work mainly utilizes the correlation between multimodal data for semantic coding. 
A multi-user multimodal semantic communication system in \cite{xie2022task}  extracts semantics for Visual Question Answersing (VQA) tasks. The multimodal semantic commucation system based on knowledge graphs is proposed in \cite{xing2024representation}, which shows channel robustness for VQA tasks. A distributed semantic communication system for audio visual parsing (AVP) tasks in \cite{wang2024distributed} reduces visual semantics transmission through audio semantic guidance. These proposed systems are oriented towards multimodal tasks. In addition, the correlation between multimodal data is used to further compress the semantics of transmission. 
A multimodal semantic communication system in \cite{li2022cross} is designed for video, audio, and tactile signals. The sharing and private semantic information is considered in \cite{gu2023semantic} for image and text, which reduces the cost of image transmission. To perform cross-modal alignment of semantics,  a semantic communication system with cross-modal alignment (CA\_DeepSC) in \cite{wang2023ca_deepsc}
uses the correlation between semantics of text and images. For multimodal tasks, a unified semantic communication system (UdeepSC) for multiple tasks in \cite{zhang2024unified} has the same performance as a single-modal system.

Although the aforementioned multimodal semantic communication systems utilize the correlation between modalities, time and semantic domain synchronization has not yet been considered. These multimodal semantic communication systems are hard to synchronize different modal features for the following reasons: 
(i) the semantic features of coding methods of each modality are usually designed independently and weakly coupled, making it hart to align when decoding semantic frames at the receiver side;
(ii) most systems are designed at the physical layer, and it is hard to match the synchronization protocols in the practice network layer. Generally, most of the synchronous control is based on data packets. In scenarios such as video conferences and live streaming, strict synchronization of audio and video is required. Therefore, synchronous semantic communication systems at higher layers are crucial for the transmission of multimodal data.

In this study, we examine a communication scenario for video conferencing, where talking-face videos and speech are transmitted synchronously. With the emergence of applications such as the burgeoning metaverse and immersive dialogues, talking-face video transmission is consuming increasing bandwidth. In \cite{jiang2022wireless}, a semantic communication system has been proposed to transmit keypoints from facial images and utilizes Hybrid Automatic Repeat reQuest (HARQ) to counteract channel variations. In \cite{tandon2022txt2vid}, speech recognition and synthesis are used to generate images synchronized with mouth movements, requiring only the transmission of text. However, these systems have not simultaneously considered the semantics of transmitting both video and speech, and the reliance on HARQ introduces additional latency. Therefore, a semantic communication system that simultaneously transmits the semantics of both data modalities is needed.
\begin{figure*}[htbp]
\centerline{\includegraphics[width=0.75\textwidth]{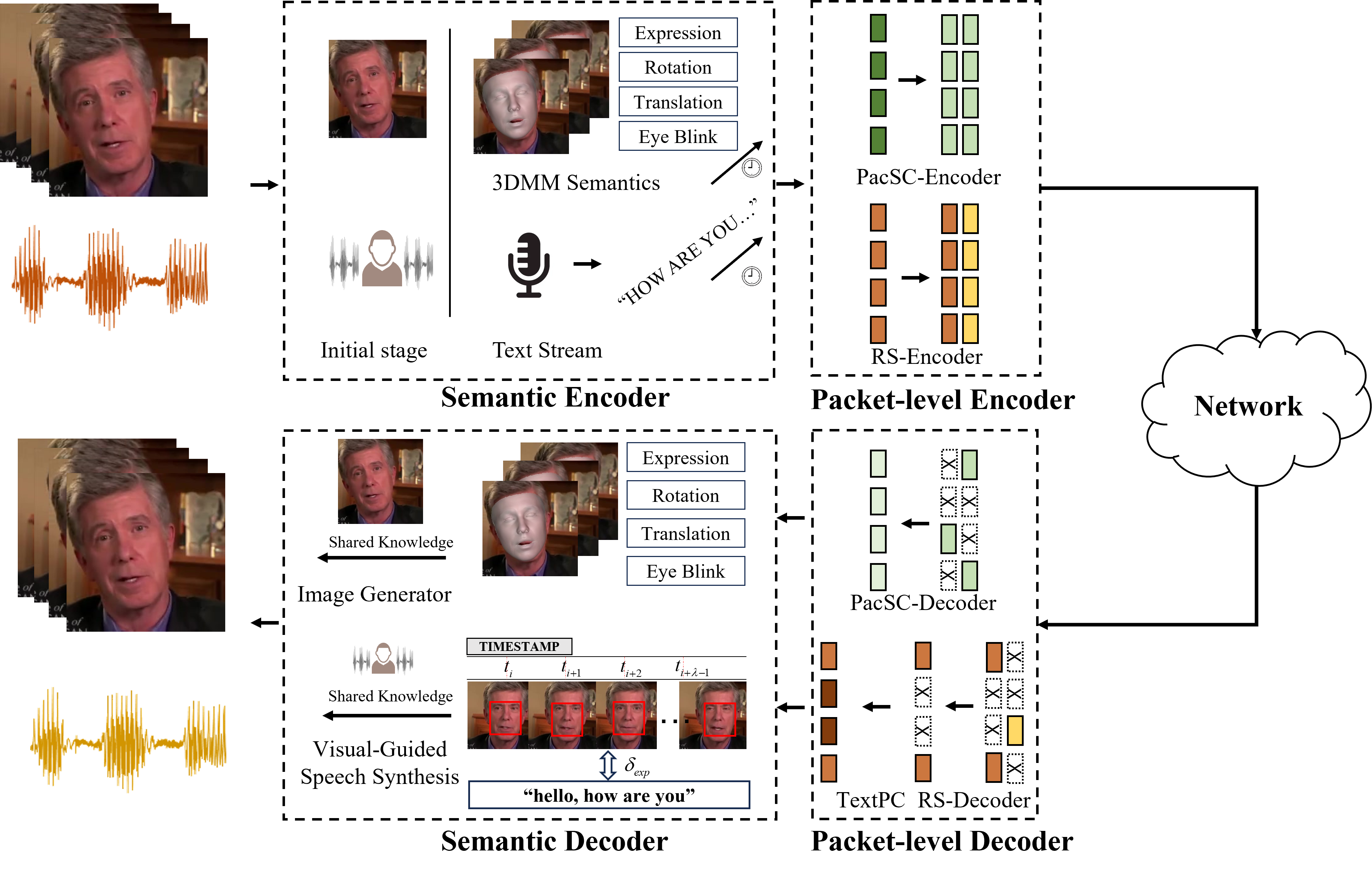}}
  \captionsetup{justification=centering} 

 \caption{The framework of synchronous semantic communication system with video-speech transmission.}
\label{fig1}
\end{figure*}
\subsection{Motivation and Contributions}
Future 6G wireless networks are required to support immersive intelligent services \cite{saad2019vision} in scenarios such as video conferencing and Extended Reality (XR), it is important to synchronous transmission of multimodal data.  Although numerous works have focused on multimodal semantic transmission, the synchronization between different modalities in terms of both the time and semantics domains has not been well addressed. Furthermore, considering that semantics are transmitted in the form of packets across networks, semantic communication systems based on physical layer designs at the symbol level are hard to synchronize. In response to network variations, packets are either perfectly received or dropped. Consequently, we consider Packet-Level Forward Error Correction (FEC) based on semantic context redundancy to address packet loss network, thereby enhancing the system's robustness.

Therefore, a synchronous semantic communication system (SyncSC) for video-speech transmission is proposed. To more effectively compress the transmission bandwidth, the transmitter utilizes 3D Morphable Model (3DMM) coefficients as the semantics for facial representation and text from speech recognition as the semantics for speech.
Initially, a reference image and a speech clip are transmitted as shared knowledge to guide the user's appearance and speech tone. The 3DMM coefficients effectively characterize facial movements and expressions, while the text from speech recognition efficiently represents the content. To maintain time synchronization, video semantics and text are time-stamped before being assembled into data packets. The receiver then synchronizes these packets on the basis of the Real-time Transport Protocol (RTP). For synchronization in the semantic domain, we design a visual-guided speech synthesis module at the receiver, which decodes synchronized speech and facial images. At the receiver, the facial semantics and text are aligned through an attention mechanism to obtain latent features, which are then input into a voice generator to produce speech with the same duration. Additionally, to combat packet loss, we design a packet-level semantic coding module 
based on Masked Autoencoders (MAE), which applies erasure coding to video semantics. Considering that the bandwidth occupied by text is already minimal, we further design a packet loss cancellation module based on pre-trained large language models to predict text content, aiming to supplement semantically similar content. Simulation results indicate that our proposed method outperforms traditional approaches in terms of performance and reduces transmission overhead, demonstrating robustness in packet loss networks.

The main contributions of this paper can be concluded as follows:
\begin{enumerate}
\item To the best of our knowledge, we are the first to propose a multimodal synchronous semantic communication system with the design of semantic packet-level FEC method above the physical layer. The semantics extracted by the transmitter remain synchronized in both the time and semantic domains and robust over packet loss network at the receiver.
\item For synchronization in the semantic domain and time domain, we design a visual-guided speech synthesis module with timestamp which uses received text and facial semantics to guide the speech synthesis.
\item We develop a semantic packet-level coding method, called PacSC, for video semantic packets based on masked autoencoders to cope with the packet loss. By leveraging semantic redundancy, our method performs better than traditional Reed-Solomon (RS) code when the network packet loss rate is high.
\item We develop a packet loss concealment module, called TextPC, for text packets based on pre-trained language models. When the packet loss rate is above 0.5, compared with RS with $\frac{1}{2}$ coderate, this module brings about a gain of 0.1 in Bilingual Evaluation Understudy (BLEU) and Semantic Similarity. 

\end{enumerate}

\subsection{Organization}
The rest of this paper is organized as follows. The framework of multimodal synchronous semantic communication system (SyncSC) is introduced in Section II. 
The details of the SyncSC are given in Section III. The simulation results are given in Section IV, and Section VI provides a conclusion.

\section{Framework of Multimodal Synchronous Semantic Communications}
In this section, the framework of the proposed system model is introduced. The system model consists of a video-speech transmitter and a multi-modal receiver.
\subsection{Video-Speech Transmitter}
As shown in Fig. 1, the proposed transmitter consists of the video semantic encoder and the speech semantic encoder. The facial semantics are extracted by the video semantic encoder, while the speech semantic encoder converts speech to text.
\subsubsection{Video Semantic Encoder} The input facial video consists of a series of frames. In particular, the input video $\boldsymbol{P} =[ \boldsymbol{p}_0,\boldsymbol{p}_1,...,\boldsymbol{p}_n,...,\boldsymbol{p}_{N_v}] $, where $N_v$ is the total number of video frames, and $\boldsymbol{p}_n\in\mathbb{R}^{256\times256\times3}$ is the $n$-th frame in RGB format. The user's appearance information serves as shared knowledge between the transmitter and receiver. The first frame $\boldsymbol{p}_0$ as the key frame is sent at the initial stage. Meanwhile, the video semantic encoder extracts the semantics of the other frames $\{\boldsymbol{p}_n,n>0\}$. Therefore, the encoded facial semantics of $\boldsymbol{p}_n$ can be expressed as
\begin{equation}
    \label{eq-2-1}
    \boldsymbol{\delta}_n ={ \boldsymbol{\rm T^V_{\alpha_v}}}(\boldsymbol{p}_n),
\end{equation}
where $ \boldsymbol{\rm T^V_{\alpha_v}}(\cdot)$ indicates the video semantic encoder with respect to parameters $\boldsymbol{\rm \alpha_v}$. The output $\boldsymbol{\delta}_n$ is a collection of 3D Morphable Mode (3DMM) \cite{3dmm} coefficients, which is a subset of the projection coefficient of the face in a three-dimensional space, expressed by
\begin{equation}
 \label{eq-2-3}
   \boldsymbol{\delta}_n = \{ \boldsymbol{\delta}_{exp_n},\boldsymbol{\delta}_{eye_n},\boldsymbol{\delta}_{rot_n},\boldsymbol{\delta}_{trans_n},\boldsymbol{\delta}_{crop_n}\},
\end{equation}
where  $\boldsymbol{\delta}_{exp_n}\in\mathbb{R}^6$ indicates the facial expression coefficients, $\boldsymbol{\delta}_{eye_n}\in\mathbb{R}^1$ indicates the eye motion coefficients, $\boldsymbol{\delta}_{rot_n}\in\mathbb{R}^3$ indicates the head rotation coefficients, $\boldsymbol{\delta}_{trans_n}\in\mathbb{R}^3$ indicates the head translation coefficients, and $\boldsymbol{\delta}_{crop_n}\in\mathbb{R}^3$ indicates the crop parameters of 3DMM. 

These coefficients are extracted directly by the pre-trained face reconstruction model \cite{deep3d}, and the parameters are not need to be finetuned. The actions and shape of the speaker's face in our semantic encoder correspond to these coefficients. Therefore, unlike existing keypoint-based semantic encoders, explicit semantic knowledge has advantages over implicit semantic features in terms of semantic editing and understanding.
\subsubsection{Speech Semantic Encoder} In order to reduce the volume of speech transmission data as much as possible, following the process in Txt2vid \cite{tandon2022txt2vid}, speech recognition is used to convert speech into text. The original input speech $\boldsymbol{S} = [\boldsymbol{s}_0,\boldsymbol{s}_1,...,\boldsymbol{s}_n,...\boldsymbol{s}_{N_s}] $ is divided into ${N_s}$ fragments and $\boldsymbol{s}_n$ is fed to the speech semantic encoder. 
In order to obtain the knowledge base of the speaker's speech style,
the first ${N_0}$ fragments are shared between the transmitter and receiver as the user's timber knowledge. The text content of the speech sample $\boldsymbol{s}_n$ can be expressed by
\begin{equation}
    \label{eq-2-4}
    \boldsymbol{l}_n = { \boldsymbol{\rm T^S_{\alpha_s}}(\boldsymbol{s}_n)},
\end{equation}
where $ \boldsymbol{\rm T^S_{\alpha_s}}(\cdot)$ indicates the speech semantic encoder with respect to the parameters $\boldsymbol{\alpha_s}$.

\subsubsection{Video Semantic Packet Encoder}
Due to the similarity in user behavior between adjacent video frames, there is semantic redundancy because of intra-frame semantic encoding. Therefore, semantic redundancy is used in the video semantic packet FEC. 
As shown in Fig. 1, 
we define the semantics of an image as the encoding symbol, and a coding block consists of $N_{p_v}$ symbols.
The output of the encoding block can be expressed by
\begin{equation}
    \boldsymbol{E}_i = \boldsymbol{\rm T^C_{\beta_v}} (\boldsymbol{\delta}_i,\boldsymbol{\delta}_{i+1},...,\boldsymbol{\delta}_{i+N_{p_v}-1}),
\end{equation}
where $\boldsymbol{E}_i = [\boldsymbol{e}_i,\boldsymbol{e}_{i+1},...,\boldsymbol{e}_{i+N_{p_v}-1}]$ indicates the output features, and $\boldsymbol{\rm T^C_{\beta_v}}$ indicates the video semantic packet encoder with respect to the parameters $\boldsymbol{\beta_v}$.

Subsequently, the semantic data packets and text with timestamps are transmitted separately to the network via the RTP protocol.

\subsection{Multi-modal Receiver}

At the receiver, arrival semantic packets are synchronized in the time domain based on their sequence number and timestamp. Our proposed multi-modal receiver first decodes the lost video packets and predicts the missing words of text. After that, the reconstructed semantics are used to generate video with speech through the Image Generator and Visual-Guided Speech Synthesis. 

\subsubsection{Video Semantic Packet Decoder}
Within the decoding block, the packet decoder decodes partially lost data packets and reconstructs semantically consistent video semantic packets. Therefore, the reconstructed semantics of the video can be expressed by
\begin{equation}
    \hat{\boldsymbol{\delta}}_i,\hat{\boldsymbol{\delta}}_{i+1},...,\hat{\boldsymbol{\delta}}_{i+N_{p_v}-1} = \boldsymbol{\rm R^C_{\theta_v}}(\tilde{\boldsymbol{e}}_i,\tilde{\boldsymbol{e}}_{i+1},...,\tilde{\boldsymbol{e}}_{i+N_{p_v}-1}),
\end{equation}
where $\boldsymbol{\rm R^C_{\theta_v}}$ indicates the video semantic packet decoder with respect to the parameters $\boldsymbol{\theta_v}$ and $\tilde{\boldsymbol{\boldsymbol{e}}}_k $, $i\leq k \leq i+N_{p_v}-1$ indicates that $\boldsymbol{\boldsymbol{e}}_k$ may be lost with a $p 
 $, $0<p<1$ probability.
\subsubsection{Text Packet-loss Concealment Module}
The existing deep learning based semantic-channel coding methods for text, such as DeepSC \cite{xie2021deep}, indicate that utilizing text redundancy requires sacrificing significant bandwidth. For example, an English word is usually represented by embedded vectors ranging from 32 to 128 dimensions, but traditional methods, such as ASCII code, only require a few bytes. Therefore, we use the traditional text encoding methods and design a packet loss concealment method based on BERT \cite{devlin2018bert} to predict the lost part at the receiver.
\begin{equation}
    \hat{\boldsymbol{L}}_i = \boldsymbol{\rm R^C_{\theta_s}}(\tilde{\boldsymbol{l}}_i,\tilde{\boldsymbol{l}}_{i+1},...,\tilde{\boldsymbol{l}}_{i+N_{p_s}-1}),
\end{equation}
where $\hat{\boldsymbol{L}}_i = [\hat{\boldsymbol{l}}_i,\hat{l}_{i+1},...,\hat{\boldsymbol{l}}_{i+N_{p_s}-1}]$ indicates the reconstructed text, $\boldsymbol{\rm R^C_{\theta_s}}$ indicates the text packet-loss concealment module with respect to the parameters $\boldsymbol{\theta_s}$, and $\tilde{\boldsymbol{\boldsymbol{l}}}_k $, $i\leq k \leq i+N_{p_s}-1$ indicates that $\boldsymbol{\boldsymbol{l}}_k$ may be lost with a $p$, $0<p<1$ probability.

\subsubsection{Image Generator}
After obtaining the decoded semantic data packet, each block $\hat{\boldsymbol{\delta}}_i$ corresponds to an image frame through the image generator. The reconstructed image $\hat{\boldsymbol{p}_i}$ can be expressed as
\begin{equation}
    \hat{\boldsymbol{p}_i} =  \boldsymbol{\rm{ R}^V_{\xi_v}}(\hat{\boldsymbol{\delta}}_i,\boldsymbol{p}_0),
\end{equation}
where $\boldsymbol{p}_0$ is the shared image between transmitter and receiver, and ${\boldsymbol{\rm{ R}^V_{\xi_v}}(\cdot)}$ indicates the image generator with parameters $\boldsymbol{\rm{\xi_v}}$.
\subsubsection{Visual-guided speech synthesis}
At the receiving end, a word corresponds to a certain length of speech and several image frames. Therefore, to align the features of the video and speech, we define a text $\boldsymbol{l}_k$ with the same timestamp $t_k$ as $\lambda$ video frames, where $\lambda$ is a positive integer, and its value depends on the block size setting of the speech recognizer and the video frame rate. Specifically, the video frames $\{\boldsymbol{p}_{\lambda k},\boldsymbol{p}_{\lambda k+1},...,\boldsymbol{p}_{\lambda (k+1)-1} \}$ and the text $\boldsymbol{l}_k$ are synchronized with the same timestamp $t_k$. The visual-guided speech synthesis module generates lip-synchronized speech $\hat{\boldsymbol{s_k}}$, by the following
\begin{equation}
  \label{eq-2-9}
    \hat{\boldsymbol{s_k}} = \boldsymbol{\rm{ R^S_{\xi_s}}}(\hat{\boldsymbol{l}}^{t_k}_s,\hat{\boldsymbol{\delta}}_{exp_{\lambda k}},\hat{\boldsymbol{\delta}}_{exp_{\lambda k+1}},...,\hat{\boldsymbol{\delta}}_{exp_{\lambda (k+1)-1}}),
\end{equation}
where $\boldsymbol{\rm{ R^S_{\xi_s}}}$ indicates the visual-guided speech synthesis with parameter $\boldsymbol{\xi_s}$. Therefore, the synchronized speech and video 
$\{ \hat{\boldsymbol{s_k}}, \boldsymbol{\hat{p}}_{\lambda k},\boldsymbol{\hat{p}}_{\lambda k+1},...,\boldsymbol{\hat{p}}_{\lambda (k+1)-1} \}$
are received.
\section{Architecture of The Proposed System}

In this section, the details of each modules in the proposed system are given. Specifically, we utilize existing deep network models to extract facial 3DMM coefficients and speech recognition for source semantic encoding. We propose MAE-based \cite{he2022masked} packet-level FEC and BERT-based \cite{devlin2018bert} packet loss concealment methods for video and speech packets to cope with the erasure channel. Moreover, at the receiver, we change the existing map-warp-edit model \cite{ren2021pirenderer}, and propose a speech synthesis module guided by facial expression coefficients.
\subsection{Semantic Encoder}

The goal of the video semantic encoder is to convert the original facial frame to accurate representations of motion. In the proposed system, we employ a subset of 3DMM parameters to represent the motion semantics. In 3DMM, the 3D shape S of a face is given by
\begin{equation}
    \label{eq-3-1}
    \boldsymbol{S} = \boldsymbol{\overline{S}} + \boldsymbol{\delta}_{id}\boldsymbol{S}_{id} + \boldsymbol{\delta}_{exp}\boldsymbol{S}_{exp},
\end{equation}
where $\boldsymbol{\overline{S}}$ indicates the average face shape. $\boldsymbol{\delta}_{id}\in \mathbb{R}^{80}$ indicates the facial shape coefficient of the identity base vector $\boldsymbol{S}_{id}$. Inspired by \cite{chen2023interactive}, $\boldsymbol{\delta}_{exp}\in \mathbb{R}^6$ is the first six of $\boldsymbol{\delta}_{exp}\in \mathbb{R}^{64}$, indicating the coefficient of facial shape of the expression base vector $\boldsymbol{S}_{exp}$. For pose semantics, head rotation is expressed as $\mathbf{r}\in SO(3)$ and translation is expressed as $\mathbf{t}\in \mathbb{R}^3$. Obtaining the 3DMM coefficient of a facial image is a regression problem. The existing 3D face reconstruction model \cite{deep3d} can extract the 3DMM coefficients of an image.
However, after experimentation, we find that $\boldsymbol{\delta}_{exp}$ is difficult to represent eye movements. Therefore, we employ the methods in \cite{eyeblink} to calculate the eye blink coefficient $\boldsymbol{\delta}_{eye}\in \mathbb{R}^1$.

For the speech semantic encoder, Zipformer \cite{yao2023zipformer}, an efficient automatic speech recognition (ASR) framework, is used to convert speech to text in real time. Specifically, the speech fragment $\boldsymbol{s}_n$ is input into the Zipformer ASR and returns the final text $\boldsymbol{l}_n$ and the timestamp $t_n$.

\subsection{Video Packet Encoder and Decoder}
As shown in Fig. 2, we propose a network based on the mask autoencoder to encode the semantics of video for the packet loss network. 
The mask autoencoder methods randomly mask partial patches at the image level and predict masked patches through adjacent patches, which is an effective method to reduce information redundancy \cite{robust}. The semantics are the projection of images in 3DMM space, so there is also significant redundancy between adjacent frames. 
Therefore, we model packet-level coding as a mask autoencoder problem and use the transformer to predict lost semantics.
The input of the encoder $X\in \mathbb{R}^{N_{p_v}\times{D_{v}}}$ consists of $N_{p_v}$ video semantics with $D_v$ dim and is divided into $M$ patches with size of $n_p\times  n_p$ through the linear projection. Following ViT \cite{dosovitskiy2020image}, linear projection is achieved through convolution and can be represented as 
\begin{equation}
    \boldsymbol{E}_p = \boldsymbol{X} \ast \boldsymbol{K} + \boldsymbol{B},
\end{equation}
where $\boldsymbol{K}$ is a convolution kernel with size of $n_p$, 
which maps the semantics to the feature space of the $D_p$ dimension, and $B$ is a $D_p$-dim bias. After adding position encoding $E_{pos}$ to the output embedded vector $E_p$, the output of linear projection can be expressed by
\begin{equation}
    \boldsymbol{E}_e = \boldsymbol{E}_p + \boldsymbol{E}_{pos},
\end{equation}
where $\boldsymbol{E}_e\in \mathbb{R}^{M \times D_p}$ and $ M = \lfloor \frac{N_{p_v}\times D_v}{n_p^2} \rfloor$. Therefore, the original video semantics are transformed into $M$ patches, fed to several Transformer Blocks \cite{vaswani2017attention}, and normalized through a Norm Layer to obtain the final encoded output $\boldsymbol{E}\in \mathbb{R}^{M \times D_p}$. 

\begin{figure}[tbp]
\centerline{\includegraphics[scale=0.6]{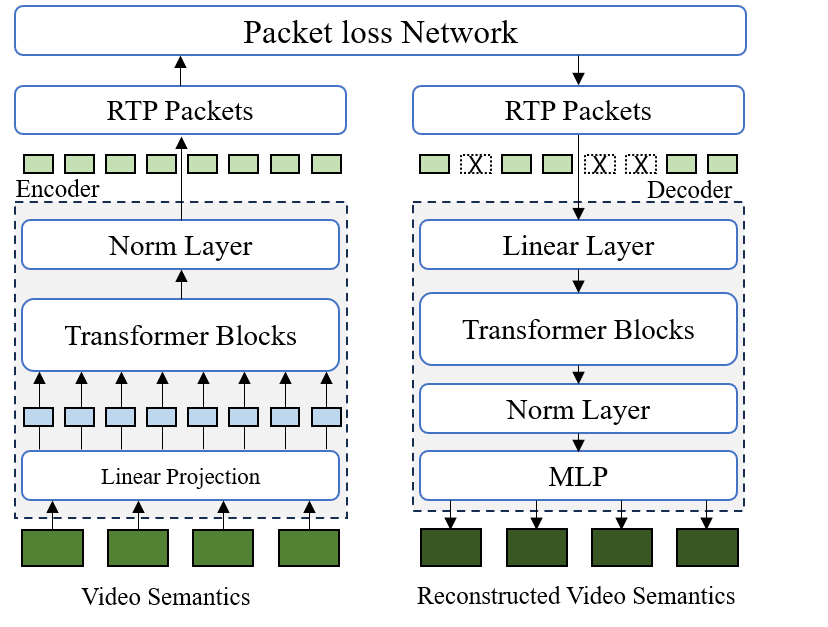}}
\caption{The architecture of video packet encoder and decoder.}
\label{Packet Encoder}
\end{figure}

The symbols of encoded semantics $\boldsymbol{E}\in \mathbb{R}^{M \times D_p}$ are divided into servral data packets 
according to the RTP protocol and sent to the network for transmission. In order to end-to-end train the packet encoder and decoder, inspired by \cite{farsad2018deep}, the dropout layer is used to simulate the network. But unlike the bit-level channels in \cite{farsad2018deep}, we use the dropout2d layer for the packet-level network. Therefore, the packet loss network can be represented as 
\begin{equation}
\label{dropout}
\boldsymbol{Y} = {\rm dropout2d} (\boldsymbol{E},p),
\end{equation}
where $\boldsymbol{Y}$ is the $N_p$ data packets observed at the receiver after RTP time synchronization, and $p$ represents the probability that a packet is dropped. 

At the receiver, we design a model with similar structure to decode partially lost packets. The linear layer is used to change the size of $\boldsymbol{Y}\in \mathbb{R}^{M \times D_p}$ to $\boldsymbol{Y}_d \in \mathbb{R}^{M \times D_d}$. Then $\boldsymbol{Y}_d$ is fed to several Transformer Blocks and a Norm Layer, $\boldsymbol{Y}_{dn}\in \mathbb{R}^{M \times D_d}$ is obtained. Finally, through a Multilayer Perceptron (MLP), the feature $\boldsymbol{Y}_{dn}$ is transformed into the size of the original input semantics $\hat{\boldsymbol{X}} \in \mathbb{R}^{N_{p_v}\times{D_{v}}}$.
\subsection{Text Packet Loss Concealment}
As shown in Fig. 3, at the transmitter, we first random interleaved encode the original text at word level. To cope with continuous packet loss,
the original text sequence, $\boldsymbol{L}=(\boldsymbol{l}_1,\boldsymbol{l}_2,...,\boldsymbol{l}_{N_{p_l}})$ is first interleaved to the sequence $\boldsymbol{L}_1=(\boldsymbol{l}^{\prime}_1,\boldsymbol{l}^{\prime}_2,...,\boldsymbol{l}^{\prime}_{N_{p_l}})$. We adopt a simple random interleaving strategy in a sending window. 
Then, the source encoding and conventional packet-level FEC, such as Huffman-RS coding, are performed on the interleaved text sequence $\boldsymbol{L}_1$ to form several RTP data packets. These data packets are sent to the packet loss network. 
At the receiver, lost words after interleaved decoding are predicted by our BERT-based model which is finetuned 
on the knowledge base between the transmiter and receiver.
\begin{figure}[tbp]
    \centerline{\includegraphics[scale=0.6]{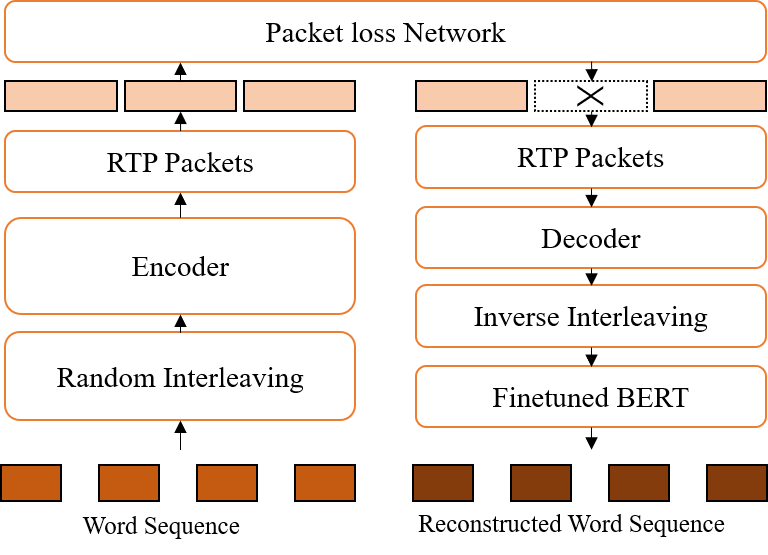}}
    \caption{The architecture of text packet loss concealment.}
    \label{fig:enter-label}
\end{figure}
\subsection{Semantic Decoder}
\begin{figure}[tbp]
\centerline{\includegraphics[scale=0.3]{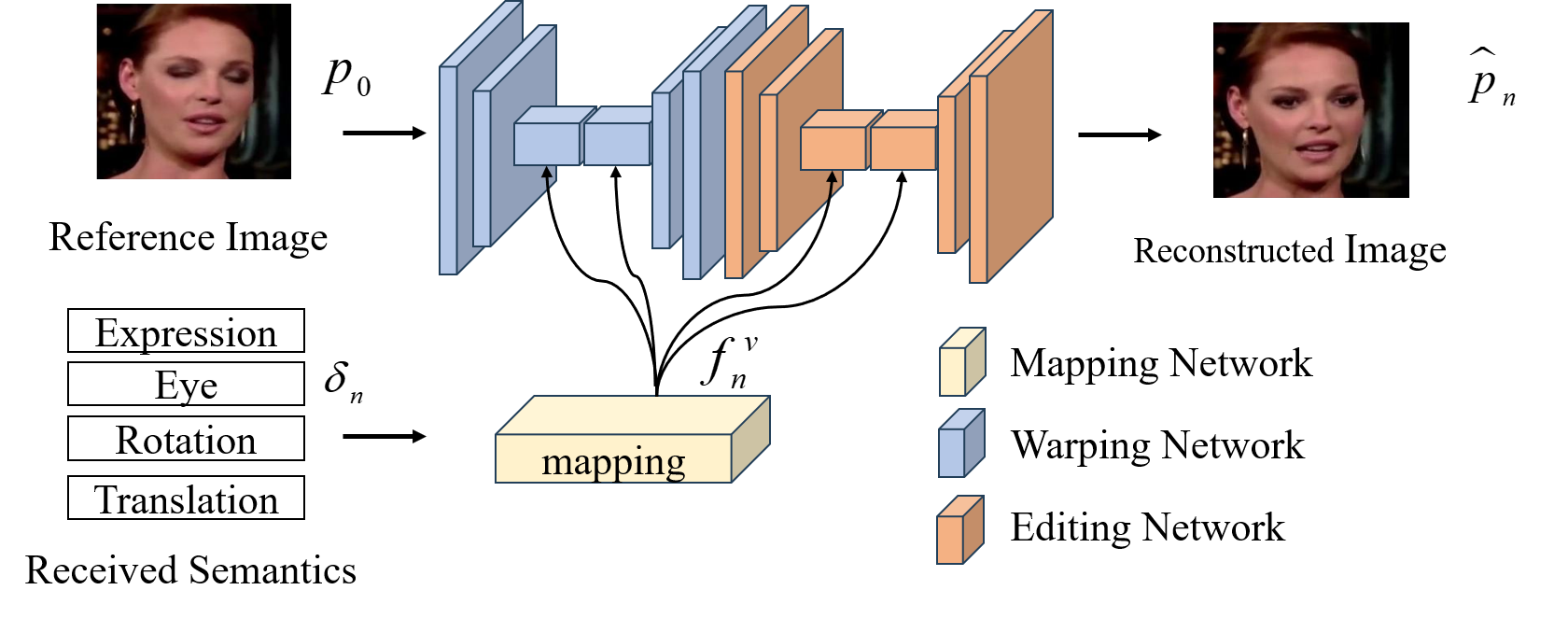}}
\caption{The architecture of image generator.}
\label{image generator}
\end{figure}
After the lost packets are decoded, 
the proposed semantic decoder synchronously generates images and speech of the speaker. The input of the semantic decoder consists of four parts: shared knowledge including reference image, reference speech samples, estimated semantics, and text. 

For decoding video semantics, as shown in Fig. \ref{image generator}, the Image Generator is made up of three networks, 
namely the mapping-warping-editing network. The reference frame $\boldsymbol{p}_0$ serves as the user's appearance feature, 
and the received semantics drive the generation of the target frames. Different from the structure in \cite{ren2021pirenderer}, the eye feature is added as input to the generator. 
Specifically, $\hat{\boldsymbol{\delta}}_n$ are first mapped to 256-dimension feature $\boldsymbol{f}_n^v$. 
Then, the reference image $\boldsymbol{p}_0$ and the mapped feature $\boldsymbol{f}_n^v$ are used to generate the flow field $\boldsymbol{w}_n$ and the warped image $\boldsymbol{\hat{p}}_{\boldsymbol{w}_n}$ through the warping network. 
The warped image $\boldsymbol{\hat{p}}_{\boldsymbol{w}_n}$ lacks details of the target image. Therefore, the reference image and the warped image are fed to the editing network to restore more details of the image $\hat{\boldsymbol{p}}_n$. 

For the speech semantic decoder, we propose a visual guided speech synthesis module. If only text content is considered, it is hard to synthesis the speech which is synchronized with video. The variation, style, and duration of the speech are difficult to estimate through text.
In the talking-face scenario, the speaker's mouth movement provides visual guidance for speech synthesis. 
As shown in Fig. 5, the proposed visual-guided speech synthesis module consists of a lip encoder, a content encoder \cite{chen2022v2c}, a style encoder \cite{chen2022v2c}, a duration aligner \cite{cong2023learning2dub}, and an audio generator. In a certain time, we employ the expression coefficient sequence of the received semantics $\hat{\boldsymbol{\zeta}}_{T_v} = [\hat{\boldsymbol{\delta}}_{exp_1},\hat{\boldsymbol{\delta}}_{exp_2},...,\hat{\boldsymbol{\delta}}_{exp_{T_v}}]$ and the text sequence $\hat{\boldsymbol{L}}_{T_l} = [\hat{\boldsymbol{l}}_1,\hat{\boldsymbol{l}}_2,...,\hat{\boldsymbol{l}}_{T_l}] $ to guide the synchronous synthesis of speech $\hat{\boldsymbol{S}}_{T_s}$, where $T_v$ is the length of received expression coefficients, $T_l$ is the number of words in the received text, and $T_s$ is the duration of the generated speech. The data of these three modalities all have the same start and end timestamps to ensure temporal alignment. Specifically, the text sequence received $\hat{\boldsymbol{L}}_{T_l}$ is encoded to a series of phoneme embeddings $\boldsymbol{\alpha}=[\boldsymbol{\alpha}_1,\boldsymbol{\alpha}_2,...,\boldsymbol{\alpha}_{T_l}]$ by the content encoder. Meanwhile, the expression coefficients sequence $\hat{\boldsymbol{\zeta}}$ is encoded to lip embeddings $\boldsymbol{E}_{lip}$ by the lip encoder. We employ a MLP to encode the expression semantics, which extracts the short-term relationship between lip motions. Then, to align the feature representations of text and facial expression, the duration aligner is used to learn the synchronous context of lip movement and phoneme. 
\begin{figure}[tbp]
    \centering
    \includegraphics[scale=0.5]{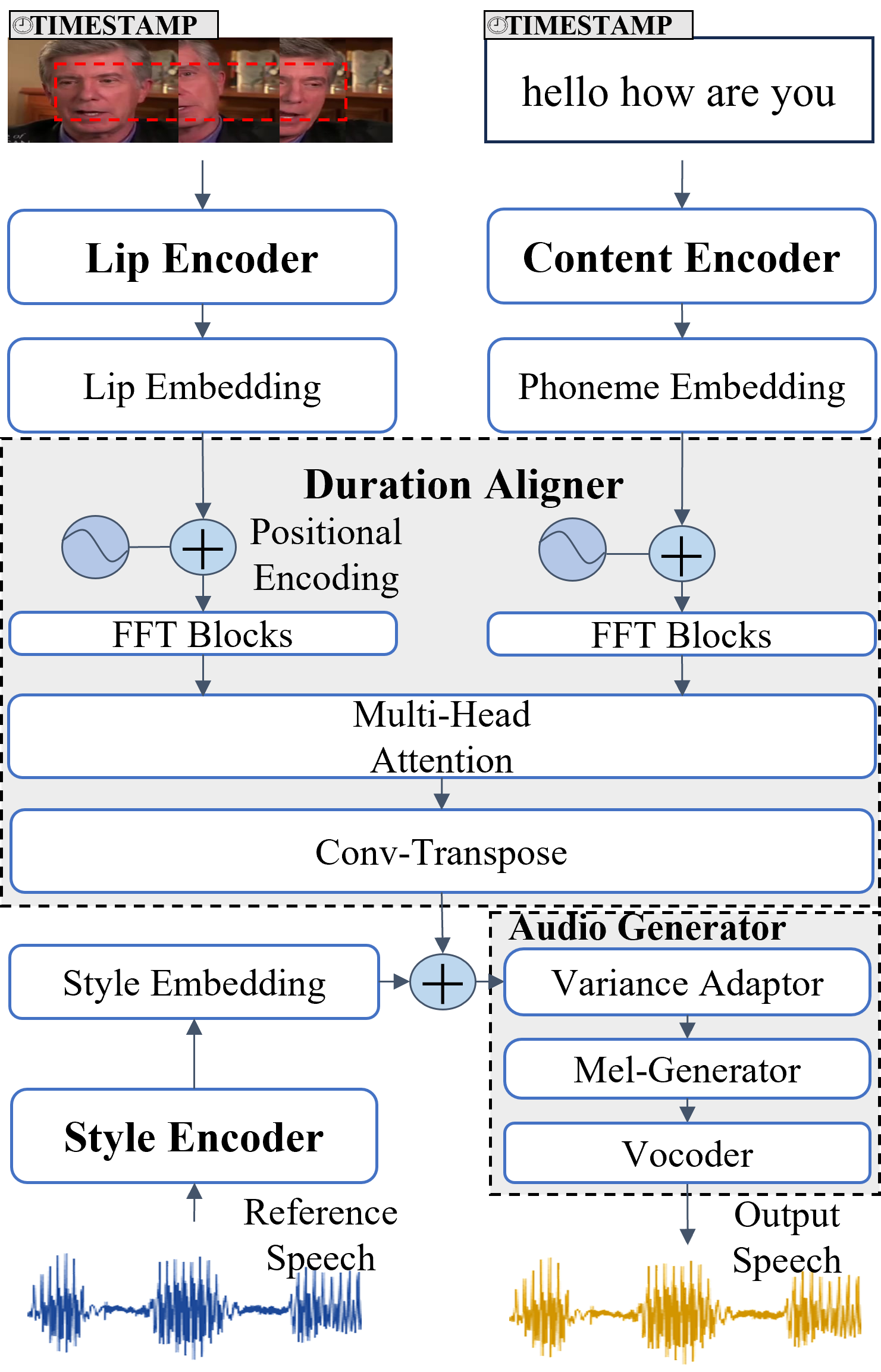}
    \caption{The architecture of visual-guided speech synthesis module.}
    \label{fig:speech decoder}
\end{figure}
After positional encoding added to lip and phoneme embeddings, Feed-Forward-Transformer (FFT) Blocks and Multi-head Attention are used to learn the context of input features. Specifically, phoneme embeddings are the query and $\boldsymbol{\alpha}$ is the key and the value of Scaled Dot-Product Attention \cite{vaswani2017attention}. Therefore, the context sequence $\boldsymbol{E}_\text{lip-text}$ of phoneme and lip embeddings is given by:
\begin{equation}
  \label{eq-3-3}
  \begin{aligned}
  \boldsymbol{E}_\text{lip-text} &= \text{Attention}(\boldsymbol{E}_{lip},\boldsymbol{\alpha},\boldsymbol{\alpha}) \\
  &=\text{Softmax}(\frac{\boldsymbol{E}_{lip}\boldsymbol{\alpha}}{\sqrt{d}})\boldsymbol{\alpha} \\
  &= \boldsymbol{A} \boldsymbol{\alpha} \in \mathbb{R}^{T_v\times d},
  \end{aligned}
\end{equation}
where  $\boldsymbol{A}$ is learnable attention weights matrix, $d$ is the dimension of context. The context sequence $\boldsymbol{E}_\text{lip-text}$ needs to be extended to the length of generated speech. According to \cite{hu2021neural}, 
if a video is synchronized with audio and picture, the length of mel-spectrograms of audio is $n$ times the frame rate of the video. The number $n$ can be expressed as
\begin{equation}
      n= \frac{T_{mel}}{T_v}=\frac{{\rm sr}/{\rm hs}}{\rm fps}\in \mathbb{N}^+,
\end{equation}
where $\rm sr$ represents the sampling rate of the speech, $\rm hs$ denotes the hop size set in calculating mel-spectrograms, and $\rm fps$ denotes the frames number per second of video. Therefore, to achieve the target length $T_{mel}=nT_{v}$, the lip-text context $\boldsymbol{E}_\text{lip-text}$ can be expanded through transposed convolutions \cite{cong2023learning2dub}. The style $\boldsymbol{E}_{\text{ref}}$ of the speaker is extracted from speech samples $\boldsymbol{s}_0$ by the style encoder. Similarly, $\boldsymbol{E}_{\text{ref}}$ is expanded to
$\boldsymbol{E}^{exp}_\text{ref}$  with length $T_{mel}$. After that, the expanded speech style feature is added to the lip-text context:
\begin{equation}
\label{eq-3-4}
\boldsymbol{E}_\text{sp} =  \boldsymbol{E}^{exp}_\text{ref} + \boldsymbol{E}^{exp}_\text{lip-text}.
\end{equation}
To enhance the details of speech synthesis and convert learned features into waveforms, the variation adaptor in \cite{ren2020fastspeech} is used to learn the pitch and energy of speech, and the mel-generator \cite{cong2023learning2dub} converts the adapted features into mel spectrograms. Speech waveforms $\hat{\boldsymbol{S}}_{T_s}$ are generated by the vocoder, HiFiGAN \cite{kong2020hifi}.
\subsection{Loss function}
Due to the fact that the trained model is used as proposed semantic encoders, further training of the parameters ${ \boldsymbol{\rm {\alpha_v}}}$ and ${ \boldsymbol{\rm {\alpha_s}}}$ is not required. The parameters of the rest of the proposed system need to be trained.
\begin{algorithm}[tb]
    \DontPrintSemicolon
     \caption{Training Algorithm for Video Semantic Packet Coding}
     \KwIn {Source Packets $X\in \mathbb{R}^{N_{p_v}\times D_v }$ from train dataset, packet loss probability $p$.}
     \KwOut {Trained Packet Encoder and Decoder model $\boldsymbol{\rm T^C_{\beta_v}}(\cdot)$ and $\boldsymbol{\rm R^C_{\theta_v}}(\cdot)$.}
       Initialization: initialize parameters $\boldsymbol{\beta_v},\boldsymbol{\theta_v}$.
    
    \While{$\mathcal{L}_{p}$ is not converged}
      {Generate semantic packets for sending: $\boldsymbol{\rm T^C_{\beta_v}}(X) \rightarrow E$.\\
      Get the received data $Y$ via (\ref{dropout}). \\
      Compute reconstructed video semantics: $\boldsymbol{\rm R^C_{\theta_v}}(Y) \rightarrow \hat{X}$. \\
      Calculate the reconstruction loss $\mathcal{L}_{\text{recon}}$ via (\ref{L_rc}).\\
      Calculate the adversarial loss $\mathcal{L}_{G}$ via (\ref{L_adv}).\\
      Calculate the total loss $\mathcal{L}_{p}$ via (\ref{Lp}).\\
      Update parameters $\boldsymbol{\beta_v},\boldsymbol{\theta_v}$ based on $\mathcal{L}_{p}$ via Stochastic Gradient Descent.\\
      Calculate the discrimintor loss $\mathcal{L}_{D}$ via (\ref{L_disc}).\\
      Update parameters of $D$ based on $\mathcal{L}_{D}$ via Adaptive Moment Estimation (Adam). \\
      Update input from the train dataset. \\
    }

\end{algorithm}
\subsubsection{Loss function of Video Semantic Packet Encoder and Decoder} The content of the reconstructed packet to be as consistent as possible with the content of the sent packet. The Huber loss is adopted as the reconstruction loss for video semantic packet coding, which can be expressed as
\begin{equation}
    \label{L_rc}
    \mathcal{L}_{\text{recon}}(X, \hat{X}) = 
    \begin{cases} 
        \frac{1}{2}(X - \hat{X})^2, & \text{if } \mid X - \hat{X} \mid \leq \gamma \\ 
        \gamma \mid X - \hat{X} \mid - \frac{1}{2}\gamma^2, & \text{if } \mid X - \hat{X} \mid > \gamma 
    \end{cases}
\end{equation}
where $\gamma$ is the the hyperparameter of the loss function. However, the Huber loss makes it difficult to learn the distribution of data within the packet. Inspired by \cite{ye2020deep}, the generative adversarial net (GAN) loss is used to learn distribution, which can be expressed by
\begin{equation}
    \label{L_adv}
    \mathcal{L}_G = \min \mathbb{E}_{Y\sim p_{Y}} [\log(1-D(G(Y))],   
\end{equation}

\begin{align}
    \label{L_disc}
    \mathcal{L}_D &= \max \mathbb{E}_{X\sim p_{data}(X)}[\log D(X)] \notag \\ 
    &+ \mathbb{E}_{Y\sim p_Y(Y)}[\log(1-D(G(Y)))],
\end{align}

where $G$ indicates the video semantic decoder, $p_{data}$ is the  distribution of original data $X$, and $p_{Y}$ is the distribution of received data $Y$ through packet loss network. We only use the Linear projection layer as the discriminator $D$ to speed up training. Therefore, the final loss function is
\begin{equation}
    \label{Lp}
    \mathcal{L}_{p} = \lambda_1 \mathcal{L}_{recon} +\lambda_2 \mathcal{L}_{G},
\end{equation}
where $\lambda_1,\lambda_2$ are the weights of the loss function, respectively. The training process is shown in Algorithm 1.
\subsubsection{Loss function of Text Packet Loss Concealment}
The prediction of text is a multi classification problem, therefore, the cross entropy function is used to train the prediction of lost words, which can be expressed as
\begin{equation}
\mathcal{L}_{l} = -\frac{1}{N} \sum_{i=1}^{N} \log(P(\boldsymbol{l}_i|\tilde{\boldsymbol{l}_i})),
\end{equation}
where $N$ indicates the number of loss words and $P(\boldsymbol{l}_i|\tilde{\boldsymbol{l}_i})$ is the conditional probability output by the model.
\begin{algorithm}[t]
\DontPrintSemicolon
  \caption{Training Algorithm for Image Generator}

  \KwIn {Source image $\boldsymbol{p}_s$ and target image $\boldsymbol{p}_t$ from train dataset $\mathcal{B}$.}
       Generate 3DMM coefficients $\boldsymbol{\delta}_s$ of $\boldsymbol{p}_s$.

  Initialization: initialize parameters $\boldsymbol{\xi}_v^0$.
  
  \While{$\mathcal{L}_{image}$ is not converged}
      {$\boldsymbol{\rm T^V_{\alpha_v}}(\boldsymbol{p}_t) \rightarrow \boldsymbol{\delta}_t $.\\      
      $\boldsymbol{\rm{ R^V_{\xi_v}}}(\boldsymbol{p}_s,\boldsymbol{\delta}_t) \rightarrow \boldsymbol{\hat{p}}_t$.\\
      Compute loss $\mathcal{L}_{image}(\boldsymbol{\xi_v})$ via (\ref{4-1-4}).\\
      Update input from the train dataset. \\
      Update parameters $\boldsymbol{\xi_v}$ via Adam.}
   
   \KwOut {Trained Image Generator $\boldsymbol{\rm{ R^V_{\xi_v}}}(\cdot)$.}
\end{algorithm}
\subsubsection{Loss function of Image Generator}
Following the loss functions used in \cite{ren2021pirenderer}, there are two stages to train the Image Generator. At the warping stage, the warping network is trained by the warping loss $L_w$, which is the perceptual loss \cite{johnson2016perceptual} between the warped image $\hat{\boldsymbol{p}}_{\boldsymbol{w}_n}$ and the target image $\boldsymbol{p}_n$, which can be expressed by
\begin{equation}
  \mathcal{L}_w = \sum_i \Vert \phi_i(\boldsymbol{\hat{p}}_{\boldsymbol{w}_n})-\phi_i(\boldsymbol{p}_n) \Vert_1,
\end{equation}
where $\phi_i$ is the activation map of the $i$-th layer of the pre-trained VGG-19 network. Then, at the editing stage, the reconstruction loss is used to measure the difference between the output image $\boldsymbol{\hat{p}}_n$ and the target image $\boldsymbol{p}_n$, which is also represented as a perceptual loss:
\begin{equation}
  \mathcal{L}_r = \sum_i \Vert \phi_i(\boldsymbol{\hat{p}}_n)-\phi_i(\boldsymbol{p}_n) \Vert_1,
\end{equation}
In addition, the statisitc error between the activation maps of VGG-19 is used to measure visual loss $\mathcal{L}_s$, which can be expressed as
\begin{equation}
\label{4-1-3}
\mathcal{L}_s = \sum_i \Vert G_i^{\phi}(\boldsymbol{\hat{p}}_n)-G_i^{\phi}({\boldsymbol{p}}_n)\Vert_1,
\end{equation}
where $G_i^{\phi}$ denotes the Gram matrix constructed from activation maps $\phi_i$. Therefore, the loss of the Image Generator is given by:
\begin{equation}
\mathcal{L}_{image} = \lambda_w\mathcal{L}_w+\lambda_r\mathcal{L}_r+\lambda_s\mathcal{L}_s.
\label{4-1-4}
\end{equation}
The image generator training algorithm is described in Algorithm 2. After obtaining the trained models of Video Semantic Packet Coder and the Image Generator, the loss function (\ref{4-1-4}) is used for end-to-end fine-tuning with $\lambda_w = 0$.
\begin{algorithm}[b]
\DontPrintSemicolon
  \caption{Training Algorithm for Visual-guided Speech Synthesis}

  \KwIn {Reference speech samples $\boldsymbol{s}_{ref}$, text sequence $\boldsymbol{W}_t$, image expression semantic sequence $\boldsymbol{\zeta_t}$ and target speech samples $\boldsymbol{s}_{t}$ from train dataset $\mathcal{D}$.}

  Initialization: initialize parameters $\boldsymbol{\xi}_s^0$.
  
  \While{$\mathcal{L}_{speech}$ is not converged}
      {$\boldsymbol{\rm{ R^S_{\xi_s}}}(\boldsymbol{s}_{ref},\boldsymbol{W}_t,\boldsymbol{\zeta}_t) \rightarrow \boldsymbol{\hat{s}}_t$.\\
      Compute loss $\mathcal{L}_{speech}(\boldsymbol{\xi_s})$ via (\ref{4-1-8}).\\
      Update input from the train dataset. \\
      Update parameters $\boldsymbol{\xi_s}$ via Adam.}
   
   \KwOut {Trained Speech Synthesis $\boldsymbol{\rm{ R^S_{\xi_s}}}(\cdot)$.}
\end{algorithm}
\subsubsection{Loss funciton of Visual-guided Speech Synthesis} 
Like the end-to-end speech synthesis in \cite{cong2023learning2dub}, the quality of the target speech and the synthesized speech is used as the loss function. Specifically, the mean square error (MSE) of pitch, energy, and mean absolute error (MAE) of a mel-spectrogram between synthesized speech and target speech are used to the loss functions:
\begin{equation}
\mathcal{L}_{pitch} = \frac{1}{T}\sum_{t=1}^{T}(\hat{P}_t-P_t)^2,
\end{equation}

\begin{equation}
\mathcal{L}_{energy} = \frac{1}{T}\sum_{t=1}^{T}(\hat{E}_t-E_t)^2,
\end{equation}

\begin{equation}
\mathcal{L}_{mel} = \frac{1}{T}\sum_{t=1}^{T}\Vert {\hat{f}}_t-f_t\Vert_1,
\end{equation}
where $P_t$, $E_t$, and $f_t$ are the pitch, energy and mel-spectrogram of $t$-th frame of target speech, and $T$ denotes the number of speech samples. Therefore, the loss of training can be expressed as:
\begin{equation}
\label{4-1-8}
\mathcal{L}_{speech} = \lambda_{p} \mathcal{L}_{pitch} + \lambda_{e} \mathcal{L}_{energy} + \lambda_{m} \mathcal{L}_{mel}.
\end{equation}
The training algorithm of Visual-guided Speech Synthesis is described in Algorithm 3.
\begin{figure*}[!tbp]
    \centering
    \begin{subfigure}{0.15\textwidth}
        \centering
        \includegraphics[width=\textwidth]{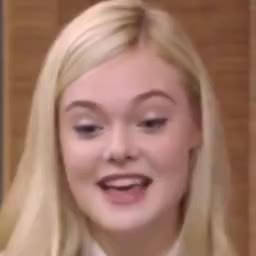}
        \caption{}
    \end{subfigure}
    \begin{subfigure}{0.15\textwidth}
        \centering
        \includegraphics[width=\textwidth]{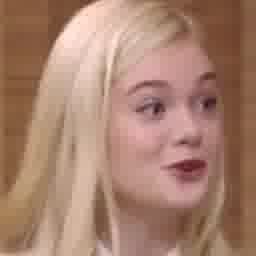}
        \caption{}
    \end{subfigure}
    \begin{subfigure}{0.15\textwidth}
        \centering
        \includegraphics[width=\textwidth]{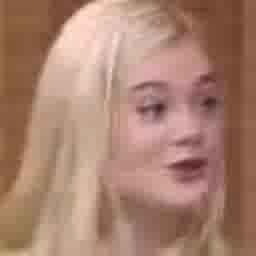}
              \caption{}
    \end{subfigure}
    \begin{subfigure}{0.15\textwidth}
        \centering
        \includegraphics[width=\textwidth]{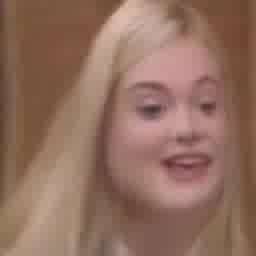}
         \caption{}
    \end{subfigure}
    \begin{subfigure}{0.15\textwidth}
        \centering
        \includegraphics[width=\textwidth]{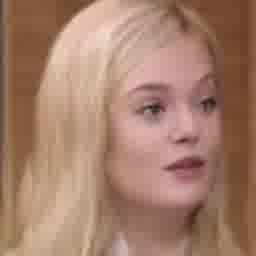}
       \caption{}
    \end{subfigure}

    \caption{Visual performance of different methods:
    (a) Original reference I-frame, (b) Original P-frame, (c) P-Frame coded by H.265 (LPIPS=0.190), (d) P-Frame coded by the FOM (LPIPS=0.243), (e) P-Frame coded by Ours (LPIPS=0.192).}
    \label{result:visual}
\end{figure*}
\begin{figure*}[!tbp]
        \centering
    \begin{subfigure}{0.15\textwidth}
        \centering
        \includegraphics[width=\textwidth]{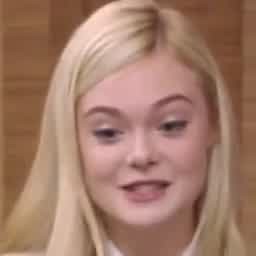}
        \caption{}
    \end{subfigure}
    \begin{subfigure}{0.15\textwidth}
        \centering
        \includegraphics[width=\textwidth]{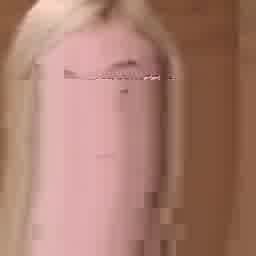}
        \caption{}
    \end{subfigure}
    \begin{subfigure}{0.15\textwidth}
        \centering
        \includegraphics[width=\textwidth]{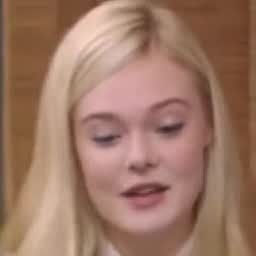}
        \caption{}
    \end{subfigure}
\caption{Visual performance through packet loss network (p=0.6): (a) Orignal Frame, (b) Received frame with H.264+RS(32,63), (c) Received frame with our Semantic-Packet Coding.}
\label{result:visual_packet}
\end{figure*}
\section{Numerical Results}
In this section, we show the numerical results of the proposed semantic system. 
\subsection{Simulation Configuration}
\subsubsection{Datasets} For training and testing video semantic coding, packet coding and image generator, we use the VoxCeleb dataset \cite{nagrani2017voxceleb} as the transmission source.  All videos in the VoxCeleb dataset are cropped to $256\times256$, which only contain facial parts. There are a total of 17913 training videos and 514 testing videos, which contain about 500 speakers. Due to the lack of alignment of speech and text in the VoxCeleb dataset, to train and test the visual-guided speech synthesis module, the Chem dataset \cite{hu2021neural} is used as the transmission source of speech and text.  We crop the videos in the Chem dataset based on subtitle files to obtain sentence-level video clips. Then, the video clips without faces are removed. There are a total of 9734 training video clips and 1500 testing sentence-level video clips. 
\subsubsection{Hyperparameter settings} 
In our proposed synchronous semantic communication system,
the code rate of video packet coding can be expressed as 
$cr = \frac{N_{p_v}\times D_v}{M \times D_p} = \frac{N_{p_v}}{\lfloor \frac{N_{p_v}\times D_v}{n_p^2} \rfloor \times D_v}$. 
Therefore, we have set different parameters $\{cr,N_{p_v},n_p,D_v\}$ for different encoding rates, such as  $\{\frac{1}{2},4,2,8\}$ and $\{\frac{1}{4},4,2,16\}$.
For text packet loss concealment, we set $N_{p_l}=6$ to balance the performance and delay. 
For the loss function, we set $\lambda_1 = \lambda_2 =0.5$, $\lambda_w=2.5$, $\lambda_r=\lambda_s=4$ and $\lambda_p=\lambda_e=\lambda_m=1$ to train the proposed model.  
\subsubsection{Baseline} We have considered end-to-end transmission scenarios in networks with a certain probability of packet loss. 
For source coding, H.264, H.265 and AV1 are used to encode video, Advanced Audio Coding (AAC), Opus are used to encode speech, and Huffman coding is used to encode the text as the conventional system.
The keypoint-based method, FOM, used in existing semantic communication systems \cite{jiang2022wireless} is the baseline for comparing video performance. Text-to-speech methods, Tacotron2 \cite{shen2018natural} and Fastspeech2 \cite{ren2020fastspeech}, of existing semantic communication systems \cite{weng2023speech} are baselines to compare speech performance. 
For packet-level forward correction of data packets, Reed solomon (RS) code in GF($2^8$) is a commonly used method in real-time video communication.

\subsection{Evaluation Metrics}
The synchronous transmission of video and speech can be described as three tasks: facial image transmission task, speech transmission task, and synchronization task. Therefore, we use the transmission quality indicators of three tasks as evaluation metrics.

\subsubsection{Facial Image Transmission Task}  
The visual quality of reconstructed images can serve as a standard for measuring transmission performance. For facial image transimission, the Structural Similarity Index Measure (SSIM) \cite{hore2010image}, the Learned Perceptual Image Patch Similarity (LPIPS) \cite{zhang2018unreasonable} and the Deep Image Structure and
Texture Similarity (DISTS) \cite{ding2020image} are utilized as metrics to measure the similarity of images between the transimter and the receiver. A lower value of LPIPS or DISTS indicates better reconstruction quality, while a higher value of SSIM indicates better quality.
\subsubsection{Speech Transmission Task} 

In this task, to compare the performance of the rate distortion with conventional methods, we use UTMOS \cite{saeki2022utmos} and Perceptual Evaluation of Speech Quality (PESQ) \cite{rix2001perceptual} as quality indicators.
UTMOS is an effective model to predict the mean opinion score (MOS) indicator to evaluate the auditory quality of audio. PESQ is one of the universal objective quality indicators. To compare the quality of text-to-speech methods, we exploit the similarity of Mel spectrograms between the reconstructed speech and ground truth as the semantic similarity. Mel Cepstral Distortion (MCD) \cite{kubichek1993mel} is utilized as the metric, which measures the distance of Mel Frequency Cepstral Coefficient (MFCC) vectors. 
A lower value of MCD represents a better quality of speech synthesis. The Structural Similarity Index (SSIM), MFCC Cosine Similarity (mfccCOS) and Fréchet Inception Distance based on MFCC (mfccFID) in \cite{li2024cm} are used to evaluate the synthesis performance of speech in various dimensions.

\subsubsection{Synchronization Task} To measure the synchronization between the reconstructed video and speech, Lip Sync Error Distance (LSE-D) and  Lip Sync Error Confidence (LSE-C) \cite{prajwal2020lip} 
are utilized as the metrics.  LSE-D calculates the average error based on the distance between the lip and speech representations. LSE-C is the confidence score that the speech and the video are synchronized with a certain time offset. Specifically, the confidence score of a synchronization error is the difference between the minimum and the median of the Euclidean distances. A lower value of LSE-D and a higher value of LSE-C represent better synchronization.

\subsection{Performance of Proposed Semantic Coding}
\begin{table}[!tbp]
    \caption{The performance versus Bpp of the proposed Video Semantic Coding and Baselines.}
    \centering
    \begin{tabular}{lccc}
    \toprule
        Method & Bpp & SSIM$\uparrow$ & LPIPS$\downarrow$  \\
        \midrule
         FOM & 0.0039 & 0.710 & 0.243 \\
         H.264 & 0.0244 & 0.877 & 0.189\\
         H.265 & 0.0119 & 0.860 & 0.190\\
         AV1 & 0.0077 & 0.867 & 0.191 \\
         \midrule
         Ours & \textbf{0.0039} & 0.745 & \textbf{0.192} \\
        \bottomrule
    \end{tabular}
    
    \label{tabel:video encoder}
\end{table}

\begin{table}
    \caption{The performance versus Bitrate of the proposed Speech Synthesis and Baselines.}
    \centering
    \begin{tabular}{lccc}
         \toprule
         Method & Bps & UTMOS$\uparrow$ & PESQ$\uparrow$\\
         \midrule
         Ground truth & 256k & 3.834 & 4.5\\
          \midrule
         AAC & 18k & 2.894 & 2.346\\
         Opus & 8k & 2.8957& 3.591\\
         \midrule
         Ours & \textbf{55.37} & 2.878 & 1.104 \\
         \bottomrule
    \end{tabular}
\end{table}
\begin{table*}
    \caption{The quality performance of the proposed Speech Synthesis and Baselines.}
    \centering
        \begin{tabular}{lcccccc}
        \toprule
Method & MCD$\downarrow$ & MCD-DTW$\downarrow$ & MCD-DTW-SL$\downarrow$ & SSIM$\uparrow$ & mfccCOS$\uparrow$ & mfccFID$\downarrow$ \\
\midrule

Tacotron2 & 15.61 & 7.92 & 11.13 & 0.13 & 0.42 & 20.16
 \\
 Fastspeech2 &17.35 &10.43 &15.32 &0.13 &0.37 &21.17\\
 Ours & \textbf{9.48} & \textbf{5.60} & \textbf{7.41} & \textbf{0.47}& \textbf{0.73}& \textbf{13.35} \\
\bottomrule
    \end{tabular}
    \label{table:speech encoder}
\end{table*}

\begin{table}
    \caption{The audio-visual synchronization performance of the proposed Speech Synthesis and Baselines.}
    \centering
    \begin{tabular}{lcc}
    \toprule
         Method & LSE-D$\downarrow$  & LSE-C$\uparrow$   \\ 
    \midrule
    Ground Truth & 8.086 & 6.127 \\
    \midrule
    Fastspeech2 & 13.052 & 2.273 \\
    Tacotron2 & 13.304 & 1.831 \\
       \midrule
       Ours & \textbf{8.822} & \textbf{5.015}\\
    \bottomrule
    \label{result:av_sync}
    \end{tabular}
\end{table}

The performance versus Bpp of the proposed video semantic coding method and baselines tested on the Voxceleb dataset are shown in Tabel. \ref{tabel:video encoder}. Our method only transmits 16 floating point numbers per frame. Similarly to the setting of the semantic communication system in \cite{jiang2022wireless}, the reference image serves as a shared knowledge base between the transmitter and receiver, and its data are not included in the calculation of the bits per pixel (bpp). We compare the FOM-based video semantic encoding methods in SVC \cite{jiang2022wireless} and set the number of keypoints to 8 and the quantization of float16 with 0.0039 bpp.
Under similar bandwidths, our method achieves an improvement of 0.035 in SSIM metrics and 0.051 in LPIPS performance in the Voxceleb dataset. 

Compared with traditional methods, the H.264 with 0.0244 bbp, H.265 with 0.0119 bbp and AV1 with 0.0077 bpp methods have similar LPIPS performance in the Voxceleb dataset.
However, the SSIM performance of traditional methods is always higher than that of our method. We present visualizations of different methods in Fig. \ref{result:visual}. Fig. \ref{result:visual}(a) and Fig. \ref{result:visual}(b) are the ground truth of the I-frame and P-frame, while Fig. \ref{result:visual}(c), Fig. \ref{result:visual}(d), and Fig. \ref{result:visual}(e) are the predicted encoding results from I-frames to P-frames. As shown in Fig. \ref{result:visual}(c), the frame coded by H.265 appear blocky visually, and our deep learning-based method has advantages in visual quality. For the FOM method, as shown in Fig. \ref{result:visual}(d) and Fig. \ref{result:visual}(e), our method has good visual performance when the character's head rotates. Therefore, our video semantic coding method can save approximately 84\% and 67\% bandwidth compared to H.264 and H.265, respectively. Moreover, compared to the AV1 method, it can save 49\% bandwidth.
\begin{figure*}[!tbp]
    \centering
    \begin{subfigure}{0.46\textwidth}
        \centering
        \includegraphics[width=\textwidth]{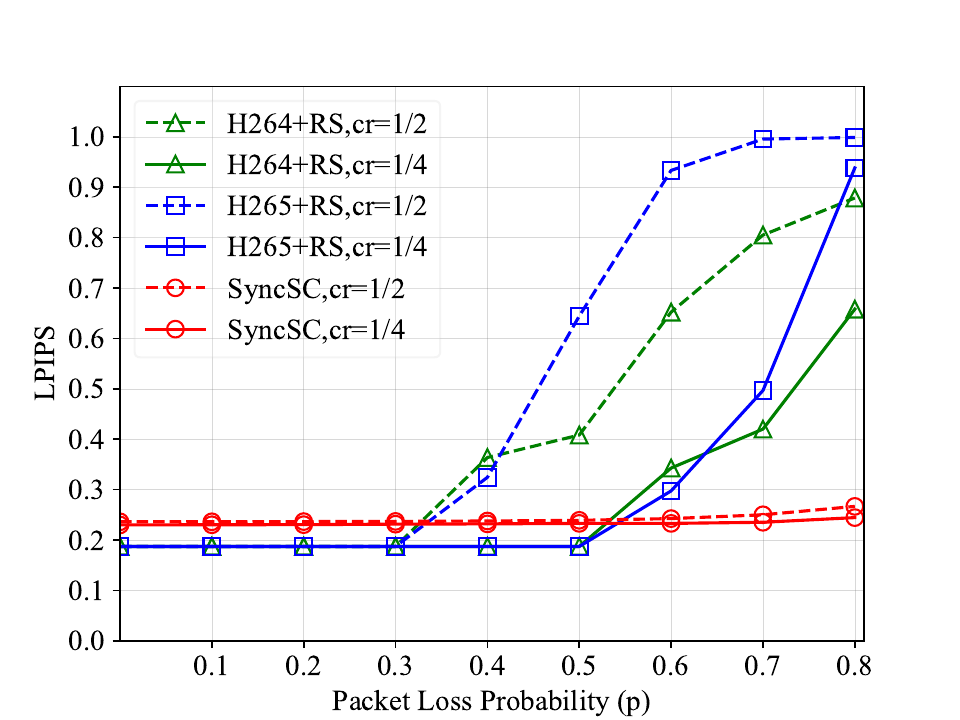}
        \caption{}
    \end{subfigure}
    \begin{subfigure}{0.46\textwidth}
        \centering
        \includegraphics[width=\textwidth]{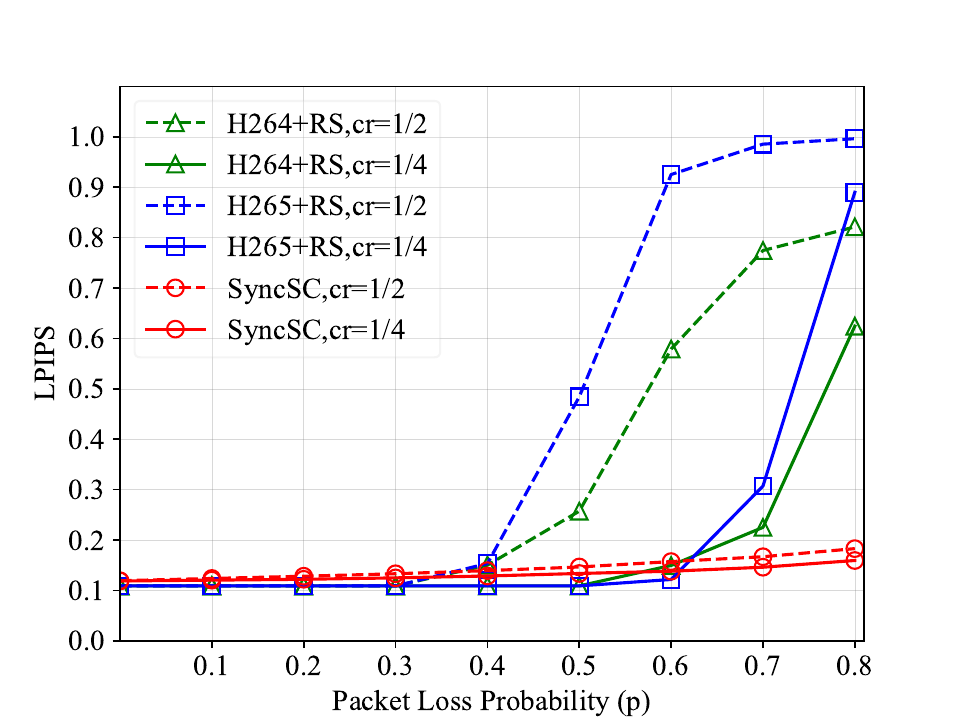}
        \caption{}
    \end{subfigure}
    \caption{LPIPS performance of the proposed system and competing methods: (a) VoxCeleb Dataset, (b) Chem Dataset.}

    \label{fig:p_ploss}
\end{figure*}

\begin{figure*}[!htbp]
    \centering
    \begin{subfigure}[b]{0.46\textwidth}
        \centering
        \includegraphics[width=\textwidth]{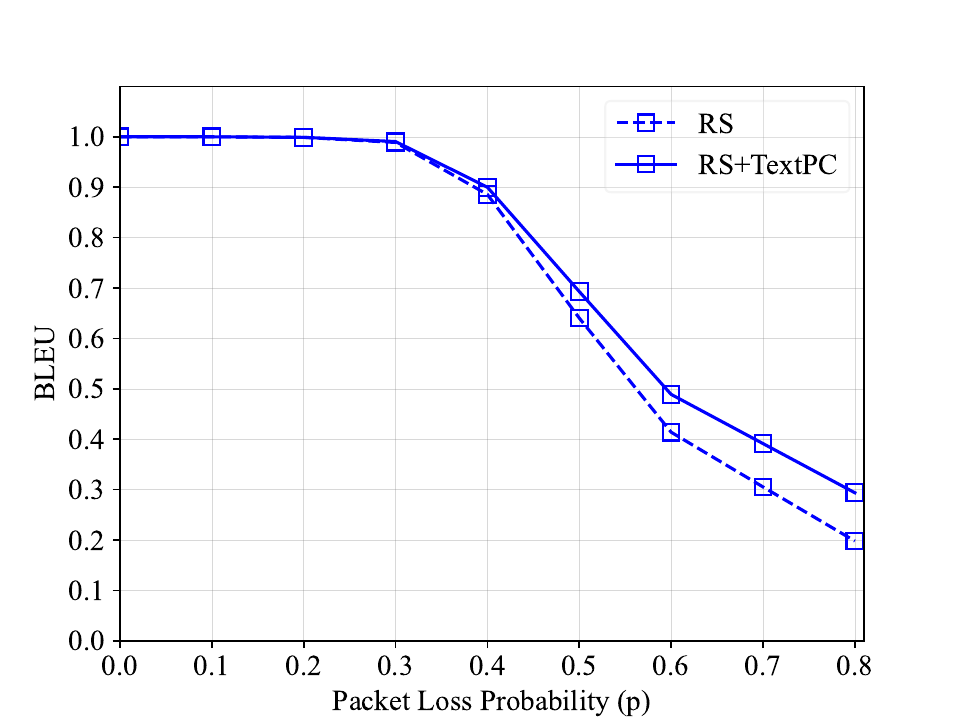}
        \caption{}
        \label{fig:sub1}
    \end{subfigure}
    \begin{subfigure}[b]{0.46\textwidth}
        \centering
        \includegraphics[width=\textwidth]{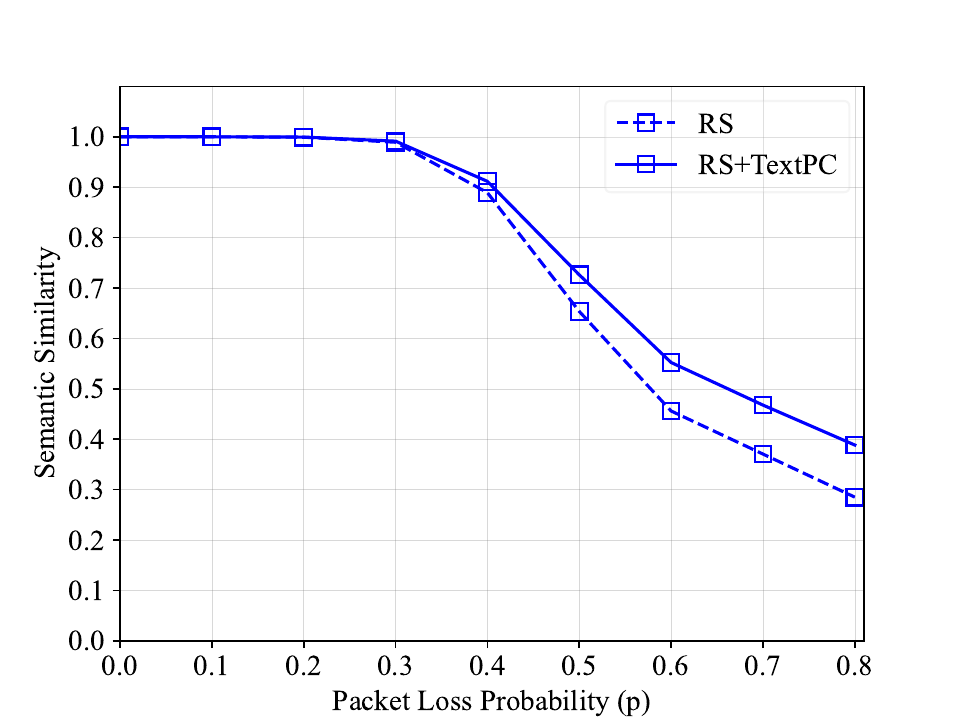}
        \caption{}
        \label{fig:sub2}
    \end{subfigure}
    \caption{(a) BLEU performance of the proposed methods. (b) Semantic similarity performance of the proposed methods with different coderates.}
    \label{fig:text_result}
\end{figure*}

The performance of the proposed speech synthesis module compared with various baselines tested on the Chem dataset is shown in Table. II and Table. III. Compared with traditional compression methods in Table. II, our text-based speech synthesis method consumes less bandwidth and achieves similar auditory quality under UTMOS indicators. However, there is a performance gap compared to AAC and Opus in the PESQ. The bit-level similarity is not guaranteed, but consistency in semantic content is considered.  In Table. III, the quality performance of speech synthesis methods used in existing speech semantic communication systems has been compared.  Our method has advantages in performance metrics. Our proposed module has significant advantages in objective indicators compared to the baseline method. Therefore, our synthesis module can significantly reduce bandwidth while exhibiting good speech quality performance.

\subsection{Performance of Synchronization}
As shown in Table. \ref{result:av_sync}, our method achieves 8.822 LSE-D and 5.015 LSE-C in terms of audio-visual synchronization performance. Compared to baselines, our method significantly improves synchronization performance by introducing visual guidance of facial information.

\subsection{Performance of Proposed Packet Coding}
The erasure channel is considered to train and test the effectiveness of our video packet-level coding method. Data packets are either perfectly received or completely lost in the network. The packet-level coding module is trained with a fixed packet loss probability, and we test the robustness with the method in networks under different packet loss probabilities.

The traditional packet-level FEC method has also been compared. We simulate the RS code in GF($2^8$) with a source length of 32 bytes based on RFC5510. H.264 and H.265 adopt the IPPP mode and encode with Group of Pictures (GOP) as an encoding block. We set the size of GOP for H.264 and H.265 to be consistent with our method at different encoding rates. Our models are trained  with a packet loss rate of 0.4 in both datasets.
Specifically, if packet loss makes H.264 or H.265 unable to decode, the LPIPS of the video is set to be 1. As shown in Fig. \ref{fig:p_ploss}, we compare the LPIPS performance of the H.264 and H.265 source encoding with RS packet-level coding under different coderates in Voxceleb and Chem dataset. For Voxceleb dataset with multiple speakers, when $cr=\frac{1}{2}$, H.264+RS and H.265+RS can repair all lost packets when $p\leq0.3$. When $cr=\frac{1}{4}$, the LPIPS performance of H.264+RS, H.265+RS and our method improve due to the introduction of more redundancy. However, when $p$ continues to increase, as shown in Fig. \ref{result:visual}(e), RS cannot reconstruct all data packets, and the LPIPS performance rapidly decays due to the emergence of mosaics. Compared to H.264, H.265 is more likely to fail to decode at high packet loss rates. Our proposed method shows a slight decrease in LPIPS performance compared to the absence of packet-level coding when $p\leq0.5$. There is still good visual performance at different packet loss rates, such as the frame shown in Fig. \ref{result:visual}(f). For Chem dataset of one speaker, H.264+RS and H.265+RS exhibits the same performance as on the Voxceleb dataset. 
Our method has good visual performance under different packet loss probabilities, even when the packet loss probability is greater than the redundancy. Therefore, our proposed method can overcome the cliff effect faced by traditional methods.

For the performance of text packet loss concealment, Fig. \ref{fig:text_result} shows the BLEU score and Semantic Similarity with respect to packet loss probability. The coderates of RS are $\frac{1}{2}$. For the BLEU score, the proposed method improves the performance of RS if $p\geq0.3$. When $p\geq 0.6$, our proposed TextPC can bring about a 0.1 improvement in BLEU. For Semantic Similarity, the proposed method significantly improves the performance of RS if $p\geq0.5$. Therefore, our proposed packet loss concealment method can predict lost words based on context at high packet loss rates without increasing bandwidth, improving the performance of BLEU and Semantic Similarity. 

\subsection{Complexity Analysis}

As shown in Table. \ref{result:complexity}, the complexity of each module in our semantic system is given. Our system is limited by the high computational complexity of video generation and speech synthesis, with end-to-end latency at second level. It is worth noting that our video semantic packet-level coder is very lightweight and has the potential to be applied in other semantic sources. In order to reduce latency, the quantization and pruning of models are considered for optimization in our future work.
\begin{table}[!tbp]
    \centering
    \caption{Complexity of the Proposed Modules.}
    \resizebox{0.48\textwidth}{!}{
    \begin{tabular}{l|ccc}
    
    \toprule
         Module & Parameters & Memory & Compute time(s) \\
         \midrule
         Video Semantic Encoder & 24.03 M & 275 MBytes & 0.041\\
         Video Semantic Decoder & 22.41 M & 318 MBytes & 0.122 \\
         Speech Semantic Encoder & - & - & 0.126 \\
         Visual-guided Speech Synthesis & 59.87 M & 648 MBytes & 0.942\\
         Video Packet Coding &  0.18 M & 825 KBytes & 0.013 \\
         Text Packet Loss Concealment & 14.20 M & 54.32 MBytes & 0.034 \\
        \bottomrule
    \end{tabular}
   }
    \label{tab:complexity}
    \begin{flushleft}
        \begin{itemize}
            \item[*] The above experimental results are conducted on the platform of Intel Xeon Gold 6342 (CPU) and one Tesla A100 (GPU).
            \item[*] The code rate of the above Video Semantic Packet Coding is $\frac{1}{2}$.
        \end{itemize}
    \end{flushleft}
    
    \label{result:complexity}
\end{table}

\section{Conclusion}
Synchronous semantic communication system for video and speech with packet-level coding is investigated.  We focus on the semantic and time domain synchronization issues in multimodal semantic communication systems, as well as the implementation of semantic packet-level forward error correction to resist erasure networks. The 3DMM coefficient is used as the transmitted semantic of video, reducing bandwidth while overcoming issues such as head turning artifacts in keypoint-based methods. The speech is converted into text for transmission, and the visual semantics of the video are used to guide the synchronous synthesis. Therefore, video and speech are synchronized in both the time and semantic domains at the receiver. Furthermore, our proposed packet-level forward error correction overcomes the cliff effect of traditional methods. Visual quality, speech quality and text semantic similarity are used to evaluate the reconstruction effect. Our proposed system can achieve lower bit rates for transmission under comparable quality of experience. At high packet loss rates, a certain level of visual and audio quality can still be guaranteed.

\ifCLASSOPTIONcaptionsoff
  \newpage
\fi



%



\bibliographystyle{IEEEtran}
\bibliography{IEEEabrv,bibtex/reference}
\end{document}